\documentclass[10pt]{IEEEtran}
\usepackage{graphicx}
\usepackage{amsmath,amssymb}
\usepackage{color}
\usepackage{epsfig}

\newcommand{\Define}{\stackrel{\triangle}{=}}

\begin{document}
\twocolumn

\title{\huge FRFD MIMO Systems: Precoded V-BLAST with Limited Feedback Versus 
Non-orthogonal STBC MIMO 
}
\author{S. Barik, Saif K. Mohammed, A. Chockalingam, and B. Sundar Rajan \\
\vspace{-0mm}
{\normalsize Department of ECE, Indian Institute of Science,
Bangalore 560012, INDIA} 
\vspace{-5mm}
}
\maketitle
\thispagestyle{empty}
\begin{abstract}
Full-rate (FR) and full-diversity (FD) are attractive features in MIMO 
systems. We refer to systems which achieve both FR and FD simultaneously 
as FRFD systems. Non-orthogonal STBCs can achieve FRFD without feedback, 
but their ML decoding complexities are high. V-BLAST without precoding 
achieves FR but not FD. FRFD can be achieved in V-BLAST through precoding 
given full channel state information at the transmitter (CSIT). However, 
with limited feedback precoding, V-BLAST achieves FD, but with some rate 
loss. Our contribution in this paper is two-fold: $i)$ we propose a 
limited feedback (LFB) precoding scheme which achieves FRFD in $2\times 2$, 
$3\times 3$ and $4\times 4$ V-BLAST systems (we refer to this scheme as 
FRFD-VBLAST-LFB scheme), and $ii)$ comparing the performances of the 
FRFD-VBLAST-LFB scheme and non-orthogonal STBCs without feedback (e.g., 
Golden code, perfect codes) under ML decoding, we show that in $2\times 2$ 
MIMO system with 4-QAM/16-QAM, FRFD-VBLAST-LFB scheme outperforms the 
Golden code by about 0.6 dB; in $3\times 3$ and $4\times 4$ MIMO systems, 
the performance of FRFD-VBLAST-LFB scheme is comparable to the performance 
of perfect codes. The FRFD-VBLAST-LFB scheme is attractive because 1) ML 
decoding becomes less complex compared to that of non-orthogonal STBCs, 
2) the number of feedback bits required to achieve the above performance 
is small, 3) in slow-fading, it is adequate to send feedback bits only 
occasionally, and 4) in most practical wireless systems feedback channel 
is often available (e.g., for adaptive modulation, rate/power control). 
\end{abstract} 

\vspace{-5.0mm}
\section{Introduction}
\label{sec1}
\vspace{-2.25mm}
Multiple-input multiple-output (MIMO) techniques can provide the benefits of
spatial diversity and multiplexing gain \cite{paulraj}-\cite{paul}. Spatial
multiplexing (V-BLAST) using $N_t$ transmit antennas achieves the full-rate 
of $N_t$ symbols per channel use. However, full transmit diversity of order 
$N_t$ is not achieved in V-BLAST. Orthogonal space-time block codes (e.g., 
$2\times 2$ Alamouti code) achieve full transmit diversity, but suffer from 
rate loss for increased number of antennas \cite{jafarkhani}. Achieving both 
full-rate (FR) and full-diversity (FD) simultaneously is a desired goal in 
MIMO communications. We refer to MIMO systems that simultaneously achieve 
both FR and FD as FRFD MIMO systems.

One way to achieve FRFD in MIMO systems is to employ non-orthogonal STBCs 
\cite{bsr}-\cite{perf07}, which offer the full-rate of $N_t$ symbols per 
channel use by having $N_t^2$ symbols in one $N_t\times N_t$ STBC matrix,
and full-diversity under ML decoding. The $2\times 2$ Golden code with 
4 symbols in one STBC matrix is a well known non-orthogonal STBC for 
2 transmit antennas \cite{gold05}. A drawback with FRFD-achieving 
non-orthogonal STBCs is their high decoding complexities, because ML 
decoding of these STBCs involves joint decoding of $N_t^2$ symbols. ML 
decoding in V-BLAST, on the other hand, involves joint decoding of only 
$N_t$ symbols. The inability of V-BLAST to achieve transmit diversity 
can be overcome through the use of precoding at the transmitter
\cite{prec1}-\cite{sumeet}. Precoding based on knowledge of full channel 
state information at the transmitter (CSIT) and first-/second-order 
statistics of the channel have been studied widely \cite{prec1}-\cite{ucd}.
With full CSIT precoding, FRFD can be achieved in V-BLAST \cite{ucd}.
However, in limited feedback precoding schemes in V-BLAST, FD is 
achieved, but with some loss in rate \cite{ASI}-\cite{sumeet}. For e.g., 
the precoding scheme in \cite{ASI} is based on Grassmannian subspace 
packing, which does not allow simultaneous transmission of more than 
$N_t-1$ streams (i.e., achievable rate is $\leq N_t-1$ symbols per 
channel use). In a $2\times 2$ MIMO system, this means a rate loss of 
50\%. The same is true with any other Grassmannian subspace packing 
based scheme or transmit antenna selection based scheme \cite{sumeet}. 
Our contribution in this paper is two fold: 
\begin{itemize}
\vspace{-1mm}
\item 
First, we present a {\em limited feedback} (LFB) based precoding scheme 
for V-BLAST  {\em which achieves FRFD} in small systems like $2\times 2$, 
$3\times 3$, and $4\times 4$ MIMO. Since the proposed scheme is not 
based on subspace packing or antenna selection, there is no loss in 
rate. The proposed scheme involves the design of a codebook having a 
finite number ($N$) of unitary precoding matrices, which are generated 
from a unitary matrix (${\bf U}_\theta$) parametrized by a single 
angular parameter, $\theta \in \{\frac{2\pi n}{N}, n=0,\cdots,N-1\}.$ 
The receiver chooses the precoding matrix which maximizes the minimum 
distance with ML decoding, and sends the corresponding index to the
transmitter. We refer to the proposed scheme as FRFD-VBLAST LFB scheme.
\item 
Second, we present a BER performance comparison between the two 
FRFD-achieving schemes, namely, the proposed FRFD-VBLAST-LFB scheme 
and non-orthogonal STBC MIMO using Golden/perfect codes under ML
decoding. Our simulation results show that in a $2\times 2$ MIMO 
system with 4-QAM/16-QAM, the proposed FRFD-VBLAST-LFB scheme 
outperforms the Golden code by about 0.6 dB. In $3\times 3$ and
$4\times 4$ MIMO, the performance of FRFD-VBLAST-LFB scheme is
comparable to the performance of perfect codes. 
\end{itemize}

The proposed FRFD-VBLAST-LFB scheme is attractive because 1) ML decoding 
becomes less complex compared to that of non-orthogonal STBCs, 2) the
 number of feedback bits required to achieve the above performance is 
small, 3) in slow-fading channels it is adequate to send 
feedback bits only occasionally, and 4) in most practical wireless 
systems feedback channel is often available (e.g., for adaptive 
modulation, rate/power control).

The rest of this paper is organized as follows. In Section \ref{sec2}, we 
present the system model. The proposed limited feedback precoding scheme
is presented in Section \ref{sec3}. BER performance of the 
proposed scheme along with a performance comparison with Golden/perfect 
codes are presented in Section \ref{sec4}. Conclusions are presented in 
Section \ref{sec5}.

\vspace{-4.0mm}
\section{System Model}
\label{sec2}
\vspace{-1mm}
Consider a precoded V-BLAST system with $N_t$ antennas at the 
transmitter and $N_r$ antennas at the receiver. Let 
${\bf H}\in \mathbb{C}^{N_r\times N_t}$ denote the channel
gain matrix, whose entries are i.i.d and $\mathcal{CN}(0,1)$.
Perfect knowledge of {\bf H} is assumed at the receiver but not
at the transmitter. 
Let $\mathcal{F} = \{{\bf F}_0,{\bf F}_1,\cdots,{\bf F}_{N-1} \}$
denote the precoder codebook of size $N$, where the ${\bf F}_n$'s,
$n=0,1,\cdots,N-1$, are $N_t\times N_t$ unitary precoding matrices. 
This codebook is known to both transmitter and receiver. For a given 
channel ${\bf H}$, the receiver chooses the precoding matrix 
from $\mathcal{F}$ that maximizes the minimum distance with ML decoding, 
and sends the corresponding index to the transmitter. Let 
$B=\lceil\log_2{N}\rceil$ denote the number of feedback bits needed to
represent this index. Given this index, $k$, the transmitter uses the 
corresponding precoding matrix, denoted by ${\bf F}={\bf F}_k$.
Let ${\bf x}\in \mathbb{A}^{N_t}$ denote the complex data 
symbol vector at the transmitter, where $\mathbb{A}$ is the modulation 
alphabet. The transmitted signal vector, ${\bf u}\in \mathbb{C}^{N_t}$
is given by ${\bf u}\, = \,{\bf F}\bf{x}$. The received signal
vector, ${\bf y} \in \mathbb{C}^{N_r}$, at the receiver is given by 
\begin{eqnarray}
{\bf y} & = & {\bf H}{\bf F}{\bf x} + {\bf n},
\label{eqn1}
\end{eqnarray}
where $n \in \mathbb{C}^{N_r}$ is the noise vector whose entries are
i.i.d $\mathcal C \mathcal N\big(0,\sigma^2=\frac{N_tE_s}{\gamma}\big)$,
where $E_s$ is the average energy of the transmitted symbols, and $\gamma$ 
is the average received SNR per receive antenna. 

\vspace{-4mm}
\section{Proposed LFB Precoding Scheme}
\label{sec3}
\begin{figure}
\begin{center}
\epsfxsize=7.00cm
\epsfysize=6.00cm
\epsfbox{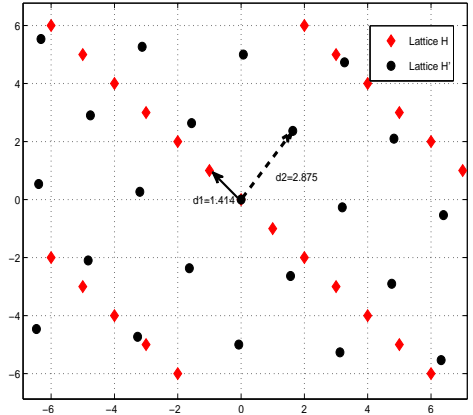}
\vspace{-2mm}
\caption{Illustration of $d_{min}$ improvement of a 2-dimensional 
lattice through transformation by a unitary matrix.}
\vspace{-7mm}
\label{fig1}
\end{center}
\end{figure}

\vspace{-2mm}
For a non-precoded system (i.e., for ${\bf F}={\bf I}_{N_t}$),
the ML decision is given by 
\begin{eqnarray}
\widehat{\bf x} &=& \arg\min_{{\bf x} \in \mathbb{A}^{N_t}} {\|{\bf y}-{\bf H}{\bf x}\|}_2^2.
\label{MLL}
\end{eqnarray}
The probability of error in the decision depends on the minimum distance,
$d_{min}$, which is given by 
\begin{eqnarray}
d_{min} & = & \min_{{\bf x}_j,{\bf x}_k \in \mathbb{A}^{N_t},\, {\bf x}_j\neq {\bf x}_k} {\|{\bf H}({\bf x}_j-{\bf x}_k)\|}_2.
\label{dmin}
\end{eqnarray}
It is known that precoding at the transmitter improves $d_{min}$ 
\cite{paul}. We illustrate this point using the following example 
and Fig. \ref{fig1}. Assume a $2\times 2$ system with {\bf H} 
{\small
$\, = \, \left[\begin{array}{cc} -1 & 5\\                                                                   1 & 3\\
\end{array}\right]$}
and PAM modulation. As can be seen from Fig. \ref{fig1}, 
$d_{min}({\bf H})=d_1=1.414$. Now, consider unitary precoding with 
{\small
${\bf F}={\left[\begin{array}{cc} \cos\frac{\pi}{6} & -\sin\frac{\pi}{6} \\
\sin\frac{\pi}{6} & \cos\frac{\pi}{6}\\
\end{array}\right]}$.
}
The new effective channel matrix is given by ${\bf H}'={\bf H}{\bf F}$.
From Fig. \ref{fig1}, it can be seen that 
$d_{min}({\bf H}')= d_2=2.875 > d_{min}({\bf H})$.

\vspace{-4mm}
\subsection{Codebook Design}
\label{sec3a}
\vspace{-1mm}
Let $\mathcal{U}_{N_t}(\theta) = \{\mbox{all } N_t\times N_t$ 
{\mbox unitary matrices parametrized by a single angular variable}, 
$\theta\}$. For e.g., for $N_t=2$, $\mathcal{U}_2(\theta)$ is the set 
of all possible matrices of the form  
$\left(\begin{array}{cc} 
A_1(\theta) & A_2(\theta)\\
A_3(\theta) & A_4(\theta) \\                                                    \end{array}\right)$, 
where $A_i(\theta)$ is a real or complex scalar function of only
$\theta$, and \\
{\small
\vspace{1mm}
$i)$ $\overline{A_1(\theta)}A_2(\theta)+\overline{A_3(\theta)}A_4(\theta)=0$,\\ 
$ii)$ $|A_1(\theta)|^2 + |A_3(\theta)|^2\ =\ |A_2(\theta)|^2 + |A_4(\theta)|^2 = 1, \ \forall \theta \in (0,2\pi).$\\
}

\vspace{-5mm}
One example of a matrix from $\mathcal{U}_2(\theta)$ is given by 
\begin{eqnarray}
{\bf U}_2(\theta) & = & \frac{1}{\sqrt{2}}\left[ \begin{array}{cc}
e^{-j\theta/2} & e^{-j\theta} \\
e^{j3\theta/2} & -e^{j\theta}
\label{eg1}
\end{array}\right]. 
\end{eqnarray}
With each matrix ${\bf U}_{N_t}(\theta) \in \mathcal{U}_{N_t}(\theta)$, 
we associate an infinite size codebook, 
{\small $\zeta_{\infty}({\bf U}_{N_t}(\theta))=\{{\bf U}_{N_t}(\theta)|_{\theta=\alpha}, \ \forall \alpha \in (0,2\pi)$\}.} 
To define a finite set precoder, we select a finite subset of size 
$N=2^B$ from $\zeta_{\infty}({\bf U}_{N_t}(\theta))$, where $B$ is the 
number of 
feedback bits. Specifically, we define the finite precoding codebook as 
$\zeta_N({\bf U}_{N_t}(\theta)) = \{{\bf U}_{N_t}(\theta)|_{\theta=\frac{2\pi i}{N}},\ 
i=0,1,\cdots,N-1)\}$. For e.g., for $N_t=2$ and $B=2$, $N=4$ with 
${\bf U}_2(\theta)$ in (\ref{eg1}), the finite precoding codebook is given by 
{\scriptsize $\zeta_4({\bf U}_2(\theta))\ =\ \left\{ \left[\begin{array}{cc}
\frac{1}{\sqrt{2}} & \frac{1}{\sqrt{2}} \\
\frac{1}{\sqrt{2}} & -\frac{1}{\sqrt{2}}\\
\end{array}\right],
\left[\begin{array}{cc}\frac{1-i}{2} & \frac{-i}{\sqrt{2}}\\
\frac{-1+i}{2} & \frac{-i}{\sqrt{2}} \end{array}\right],
\left[\begin{array}{cc}
\frac{-i}{\sqrt{2}} & \frac{-1}{\sqrt{2}} \\
\frac{-i}{\sqrt{2}} & \frac{1}{\sqrt{2}} \\
\end{array}\right], \\
\left[\begin{array}{cc}
\frac{-1-i}{2} & \frac{i}{\sqrt{2}}\\
\frac{1+i}{2} & \frac{i}{\sqrt{2}} 
\end{array}\right]\right\}$.
}
\\
The performance of the codebook is dependent on the choice of 
${\bf U}_{N_t}(\theta)$ and $N$. We, therefore, come up with the following 
performance indicator for a given codebook: 

\vspace{-6mm}
{\small
\begin{eqnarray}
\mu(\zeta_N({\bf U}_{N_t}(\theta)) & = & \\
& & \hspace{-29mm}\mathbb{E}_{\bf H} \left[\dfrac{{\mbox{max}\atop{\theta_i=\frac{2\pi i}{N}, i\in \{0,\cdots,N-1\}}} \ {\mbox{min}\atop{{\bf x}_j\neq {\bf x}_k, {\bf x}_j,{\bf x}_k \in \mathbb{A}^{N_t}}} {\big\|{\bf H}{\bf U}_{N_t}(\theta_i)({\bf x}_j-{\bf x}_k)\big\|_2^2}} {{\mbox{min}\atop{{\bf x}_j\neq {\bf x}_k, {\bf x}_j,{\bf x}_k \in \mathbb{A}^{N_t}}}{\big\|{\bf H}({\bf x}_j-{\bf x}_k)\big\|_2^2}}\right]. \nonumber
\label{metric}
\end{eqnarray}
}

In words, $\mu(\zeta_N({\bf U}_{N_t}(\theta))$ is the expected ratio of the 
maximum squared $d_{min}$ with precoding to that without precoding.
Then, the optimal finite precoding codebook with $B=\log_2N$ feedback 
bits is $\zeta_N({\bf U}_{N_t}^{opt}(\theta)$), where
${\bf U}_{N_t}^{opt}(\theta)$ is given by
\begin{eqnarray}
{\bf U}^{opt}_{N_t}(\theta) & = & {\mbox{argmax}\atop{{{\bf U}_{N_t}(\theta) \in \mathcal{U}_{N_t}(\theta)}}} \  \mu(\zeta_N({\bf U}_{N_t}(\theta)).
\label{OPEN}
\end{eqnarray}
Obtaining an exact solution for ${\bf U}^{opt}_{N_t}(\theta)$ analytically
is difficult. In the absence of a solution to the above problem, we tried
out several ${\bf U}_{N_t}(\theta)$ matrices for small values of $N_t$ (e.g., 
$N_t=2,3,4$, which are of interest in practical MIMO systems), and found 
that the following designs for $N_t=2,3,4$ work very well\footnote{Our 
computer simulations show that these designs for $N_t=2,3,4$ achieve 
very good BER performance (as we will see in Sec. \ref{sec4}).} in the 
proposed scheme:
\begin{eqnarray}
\textbf{U}_{N_t=2}(\theta)\ =\ \dfrac{1}{\sqrt{2}}\left[\begin{array}{cc}
e^{j\theta} & 1\\
-1 & e^{-j\theta}\\
\end{array}\right],
\label{2x2case}
\end{eqnarray}
\begin{eqnarray}
\textbf{U}_{N_t=3}(\theta) & = & \dfrac{1}{3}\left[\begin{array}{ccc}
2e^{j\theta} & -2 & e^{j\theta}\\
e^{j\frac{\theta}{2}} & 2e^{-j\frac{\theta}{2}} & 2e^{j\frac{\theta}{2}}\\
2 & e^{-j\theta} & -2\\
\end{array}\right],
\label{3x3case}
\end{eqnarray}
\begin{eqnarray}
\textbf{U}_{N_t=2^{m+1}}(\theta)\ =\ \dfrac{1}{\sqrt{2}}\left(\begin{array}{cc}
\textbf{U}_{2^m}(\theta) & {\bf I}_{2^m}\\
-{\bf I}_{2^m} & \textbf{U}^H_{2^m}(\theta)\\
\end{array}\right).
\label{4x4case}
\end{eqnarray}

\vspace{-3mm}
\subsection{Precoding Matrix Selection} 
\label{sec3b}
\vspace{-1mm}
At the receiver, given the knowledge of {\bf H}, we define 
$d_{min}({\bf H},i)$, $i=0,1,\cdots,N-1$, as

\vspace{-4mm}
{\scriptsize
\begin{eqnarray}
d_{min}({\bf H},i) \, \Define \, {\mbox{min}\atop{{{\bf x}_j,{\bf x}_k \in \mathbb{A}^{N_t}, {\bf x}_j\neq {\bf x}_k}}} {\Big\|{\bf H}{\bf U}_{N_t}(\theta)\big|_{\theta=\frac{2\pi i}{N}}({\bf x}_j-{\bf x}_k)\Big\|}_2. \hspace{-1mm}
\label{dmin1}
\end{eqnarray} 
}

\vspace{-3mm}
The receiver sends to the transmitter the index $p$, given by
\begin{eqnarray}
p & = & {\mbox{argmax}\atop{{i \in \{0,1,..,N-1\}}}} d_{min}({\bf H},i),
\end{eqnarray}
using $B$ bits of feedback.
Hence, the optimum precoding matrix chosen is given by
\begin{eqnarray}
{\bf F}_p & = & {\bf U}_{N_t}(\theta)\big|_{\theta=\frac{2\pi p}{N}}.
\end{eqnarray}

\vspace{-3mm}
\subsection{Receiver Processing: Feedback Computation and 
Signal Detection }
\label{sec3c}
\vspace{-1mm}
Signal detection is performed using the sphere decoding algorithm 
\cite{KAU}. In the following, we present the computation of 
$d_{min}({\bf H},i)$ in (\ref{dmin1}) for $i = 0,1, \cdots N-1$.

We can rewrite the system model equation (\ref{eqn1}) for 
the precoded system, when precoded with the $i$th precoding matrix as
\begin{equation}
\label{eq1}
{\bf y} =\ {\bf H}_i {\bf x}\ +\ {\bf n},
\end{equation}
where ${\bf H}_{i} \Define {\bf H} {\bf U}_{N_t}(\theta)\big|_{\theta=\frac{2\pi i}{N}}$.
Using $\mathcal{R}\{.\}$ and $\mathcal{I}\{.\}$ to denote the real
and imaginary parts of a complex argument,
the above equation can be transformed into an equivalent real-valued model as
\begin{equation}
\label{realeq}
\tilde{\bf y} \ = \ {\tilde {\bf H}}_i \tilde{\bf x}   \  +   \  \tilde{\bf n},
\end{equation}
where $\tilde{\bf y}\ =\ {[\mathcal{R}\{{\bf y}^T\}\ \ \ \mathcal{I}\{{\bf y}^T\}]}^T,\ \tilde{\bf x}\ =\ {[\mathcal{R}\{{\bf x}^T\}\ \  \mathcal{I}\{{\bf x}^T\}]}^T$,  \  $\tilde {\bf n}\ =\ {[\mathcal{R}\{{\bf n}^T\}\ \ \ \mathcal{I}\{{\bf n}^T\}]}^T$, and
\begin{equation}
\tilde{\bf H}_i\ =\ \left[\begin{array}{cc}
\mathcal{R}\{{\bf H}_{i}\} & -\mathcal{I}\{{\bf H}_{i}\}\\
\mathcal{I}\{{\bf H}_{i}\} & \mathcal{R}\{{\bf H}_{i}\}\\
\end{array}\right].
\end{equation}
Here, $\tilde{\bf y}\ \in\ \mathbb{R}^{2N_r\times 1},\ \tilde{\bf x}\ \in \mathbb{S}^{2N_t\times 1},\ \tilde{\bf n}\  \in\ \mathbb{R}^{2N_r\times 1}\ {\mbox and}\ \tilde{\bf H}_i \ \in \ \mathbb{R}^{2N_r \times 2N_t} $. Also, $\mathbb{S}$ is the real PAM constellation corresponding to $\mathbb{A}$.
Henceforth, we shall work with the real-valued system in (\ref{realeq}).
With the new system model in (\ref{realeq}), we can re-write (\ref{dmin1}) as

\vspace{-2mm}
{\footnotesize
\begin{eqnarray}
\label{dmin_2}
d_{min}({\bf H},i) & \hspace{-2mm} \Define & \hspace{-2mm} {\mbox{min}\atop{{{\bf x}_j,{\bf x}_k \in \mathbb{A}^{N_t}, {\bf x}_j\neq {\bf x}_k}}} {\Big\|{\bf H}{\bf U}_{N_t}(\theta)\big|_{\theta=\frac{2\pi i}{N}}({\bf x}_j-{\bf x}_k)\Big\|}_2 \nonumber  \\
& \hspace{-2mm} = & \hspace{-2mm} {\mbox{min}\atop{{\tilde{\bf x}_j,\tilde{\bf x}_k \in \mathbb{S}^{2N_t}, \tilde{\bf x}_j\neq \tilde{\bf x}_k}}} {\Big\|\tilde{\bf H}_i (\tilde{\bf x}_j-\tilde{\bf x}_k)\Big\|}_2 \nonumber \\
& \hspace{-2mm} = & \hspace{-2mm} {\mbox{min}\atop{{{\bf z} \in \mathbb{D}^{2N_t}, {\bf z} \neq {\bf 0}}}} {\Big\|\tilde{\bf H}_i {\bf z}\Big\|}_2,
\end{eqnarray}
}

\vspace{-2mm}
where $\mathbb{D} $ is the difference constellation of $\mathbb{S}$. 
For example, for square $M$-QAM modulation, $\mathbb{S}$ is given by 
$\{\pm1,\pm3,..,\pm(\sqrt{M}-1)\}$ and $\mathbb{D}$ is 
$\{0,\pm2,\pm4,..,\pm2(\sqrt{M}-1)\}$. Since the factor of 2 can be 
neglected, $\mathbb{D}$ can therefore be taken to be 
$\mathbb{D}$ = $\{s : s \in \mathbb{Z},\ |s| \ \leq (\sqrt{M}-1)\}$.

Geometrically, $\tilde{\bf H}_i$ defines a $2N_t$ dimensional lattice 
in $\mathbb{R}^{2N_r}$, denoted as 
$\Lambda\ =\ \{\tilde{\bf H}_i {\bf z} : {\bf z} \in \mathbb{Z}^{2N_t}\}$ 
and a finite subset of $\Lambda$ is 
$\tilde \Lambda$ = $\{\tilde{\bf H}_i{\bf z}:{\bf z}\in \mathbb{D}^{2N_t}\}$. 
For the receiver to find the optimal precoding matrix, it needs to evaluate 
${d}_{min}({\bf H},i)$ using (\ref{dmin_2}) for all $i=0,1, \cdots N-1$. 
Given the geometrical interpretation it is easy to see that calculation 
of (\ref{dmin_2}) is equivalent to the problem of finding the shortest 
vector in a finite subset of the lattice $\Lambda$. This problem is a 
constrained version of the well known shortest vector problem (SVP) for 
any arbitrary lattice. Unconstrained SVP can be solved by appropriately 
modifying the closest lattice point search algorithm, as discussed in 
\cite{VAR}. We solve the constrained SVP problem by restricting the 
search to be within the finite subset $\tilde \Lambda$.

\vspace{-3mm}
\section{Simulation Results}
\label{sec4}
\vspace{-1mm}
We evaluated the BER performance of the proposed limited feedback
precoding scheme as a function of average received SNR per receive
antenna, $\gamma$, through simulations for $N_t=N_r=2,3,4$. For 
comparison purposes, we also evaluate the performance of Golden 
code/perfect codes in MIMO systems with $N_t=N_r=2,3,4$.

\begin{figure}
\hspace{-4mm}
\epsfxsize=8.750cm
\epsfysize=7.25cm
\epsfbox{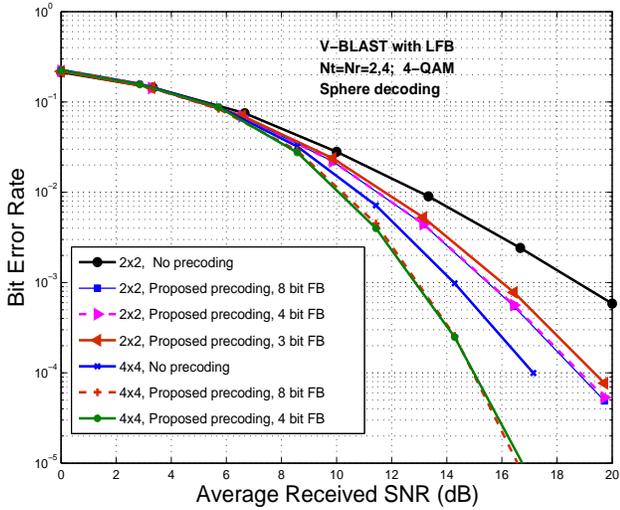}
\vspace{-2mm}
\caption{BER performance of the proposed limited feedback precoded 
$2\times 2$ and $4\times 4$ V-BLAST scheme. 4-QAM, \# feedback bits, 
$B = 8,4,3$. {\em 4-bit feedback achieves very good
performance.}}
\vspace{-4mm}
\label{fig2}
\end{figure}

\begin{figure}
\hspace{-4mm}
\epsfxsize=8.750cm
\epsfysize=7.25cm
\epsfbox{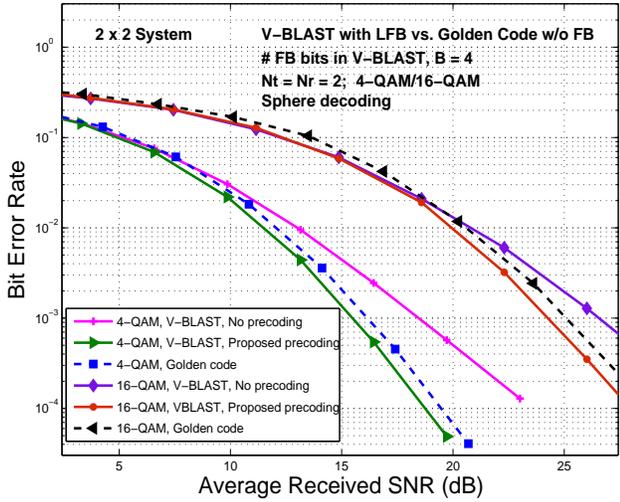}
\vspace{-2mm}
\caption{BER performance of the proposed limited feedback precoded
$2\times 2$ V-BLAST scheme versus $2 \times 2$ Golden code without
feedback. 4-QAM and 16-QAM, \# feedback bits in V-BLAST, $B = 4$.
{\em Proposed scheme achieves full-diversity; outperforms Golden 
code by about 0.6 dB.}}
\vspace{-4mm}
\label{fig3}
\end{figure}

Figure \ref{fig2} shows the simulation results for $2\times 2$ and 
$4\times 4$ V-BLAST without and with the proposed precoding, for 
4-QAM and 16-QAM using sphere decoding. BER performance plots for 
different levels of quantization requiring different number of feedback 
bits ($B=8,4,3$ bits) are shown. From Fig. \ref{fig2}, it is observed 
that the proposed precoding scheme achieves significantly better 
diversity than the `no precoding' scheme. In fact, a comparison of 
this precoded performance for $2\times 2$ V-BLAST with the performance 
of $2\times 2$ Golden code in Fig. \ref{fig3} shows that both these 
curves run parallel illustrating that, just like Golden code, the 
proposed scheme also achieves full diversity. Another interesting 
observation in Fig. \ref{fig2} is the effect of feedback bits on the 
BER performance. It can be seen that the BER with $B=4$ and $B=8$ 
are almost the same, showing that the performance in the proposed
scheme remains robust even with a nominal quantization of using 4 bits. 

\begin{figure}
\hspace{-4mm}
\epsfxsize=8.750cm
\epsfysize=7.25cm
\epsfbox{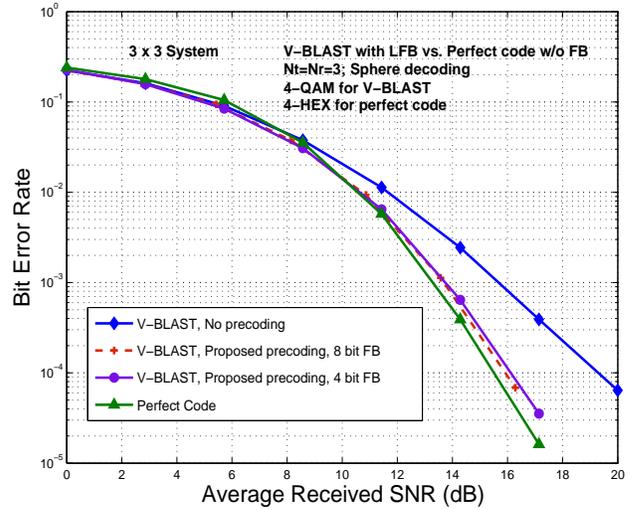}
\vspace{-2mm}
\caption{BER performance of the proposed limited feedback precoded
$3\times 3$ V-BLAST scheme versus $3 \times 3$ perfect code without
feedback. 4-QAM for V-BLAST, 4-HEX for perfect code, \# feedback bits 
in V-BLAST, $B = 8,4$. {\em Proposed scheme performs close to 
perfect code within about 0.3 dB. }}
\vspace{-4mm}
\label{fig4}
\end{figure}

\begin{figure}
\hspace{-4mm}
\epsfxsize=8.75cm
\epsfysize=7.25cm
\epsfbox{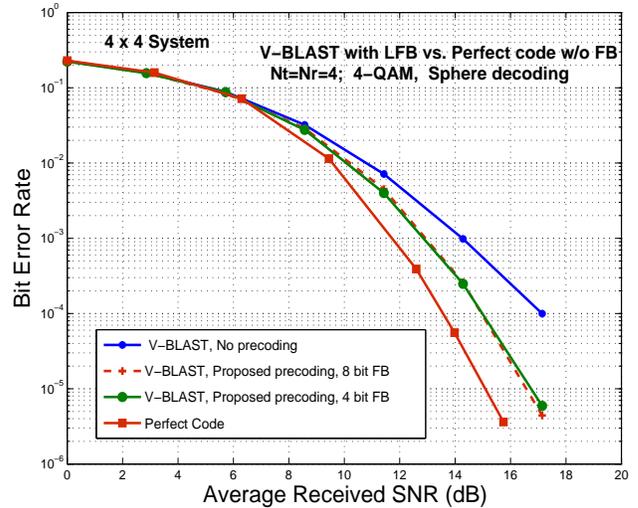}
\vspace{-2mm}
\caption{BER performance of the proposed limited feedback precoded
$4\times 4$ V-BLAST scheme versus $4 \times 4$ perfect code without
feedback. 4-QAM, \# feedback bits in V-BLAST, $B = 8,4$.
{\em Proposed scheme performs about 1 dB away from perfect code.}}
\vspace{-2mm}
\label{fig5}
\end{figure}

Next, in Figs. \ref{fig3} to \ref{fig5}, we compare the BER performances 
of the proposed precoding scheme and the Golden code/perfect codes in 
$2\times 2$ (Fig. \ref{fig3}), $3\times 3$ (Fig. \ref{fig4}), and 
$4\times 4$ (Fig. \ref{fig5}) systems, using sphere decoding.  
The channel is assumed to remain constant for $N_t$ consecutive 
channel uses in V-BLAST in order to facilitate the comparison 
between V-BLAST and Golden code/perfect codes (which are assumed
to have a quasi-static interval of $N_t$ channel uses) under 
similar quasi-static channel conditions. 

From Fig. \ref{fig3}, it can be seen that, for both 4-QAM and 16-QAM,
the performance curves of the proposed scheme runs parallel to those of 
the Golden code, showing that the proposed scheme achieves the full 
diversity of 4. It is interesting to further observe that the proposed 
scheme exhibits some coding gain advantage compared to the Golden 
code. Particularly, the coding gain attained by the proposed scheme 
over Golden Code is about 0.7 dB at a BER of $10^{-3}$. This is 
significant since precoding achieves this better performance with a 
lower decoding complexity (joint detection of 2 symbols in one channel
use) than the Golden code (joint detection of 4 symbols in one STBC
matrix).

Next, the BER comparison in Fig. \ref{fig4} for $3\times 3$ system
shows that the proposed scheme achieves the same diversity as that 
of $3\times 3$ perfect code, but is inferior to perfect code in terms 
of coding gain. This performance gap, however, is small (about 0.3 dB). 
In $4\times 4$ system in Fig. \ref{fig5}, the performance gap in terms 
of coding gain is about 1 dB. This suggests that better precoding 
strategies for larger $N_t$ can be investigated to achieve close to
perfect code performance. A likely approach can be to consider multiple
parameter based  precoder designs. 

\begin{figure}
\hspace{-4mm}
\epsfxsize=8.750cm
\epsfysize=7.25cm
\epsfbox{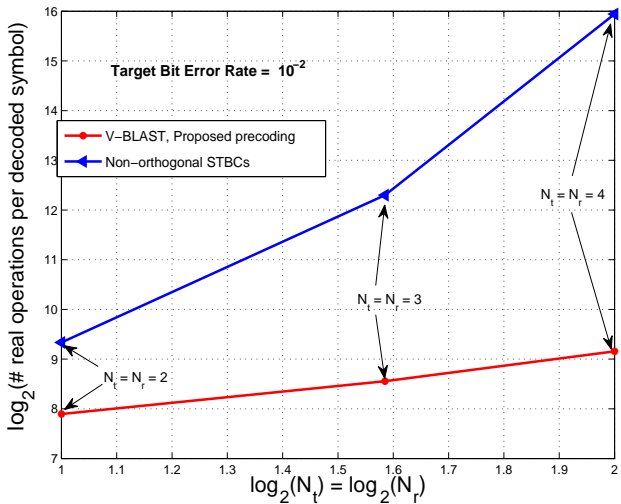}
\vspace{-2mm}
\caption{Decoding complexity (in terms of \# real operations per decoded
symbol) of the proposed scheme versus Golden code/perfect codes at SNRs
corresponding to a target BER of $10^{-2}$. 4-QAM, $N_t=N_r=2,3,4$. 
\# feedback bits in V-BLAST, $B = 4$, sphere decoding.
{\em Decoding complexity in the proposed scheme is much less than
in Golden code/perfect codes.}}
\vspace{-2mm}
\label{fig6}
\end{figure}

A key advantage of the proposed precoder approach compared to the
non-orthogonal STBC approach is its lesser decoding complexity. 
This advantage is captured in the complexity comparison plots in 
Fig. \ref{fig6}, where the complexities in both the approaches using
sphere decoding, in terms of number of real operations per decoded symbol, 
are plotted as a function of $N_t=N_r=2,3,4$ at SNRs corresponding to a 
target BER of $10^{-2}$. It can be seen that the complexity in the 
proposed approach is much less because it needs to jointly detect 
only $N_t$ symbols, whereas in the non-orthogonal STBC approach 
joint detection is over $N_t^2$ symbols. We note that the complexity 
comparison in Fig. \ref{fig6} does not include the complexity involved 
in the optimization to choose the optimum precoding matrix at the 
receiver. It is pointed out that the data decoding complexity dominates
over the precoding selection complexity, as decoding is done on a
per channel use basis whereas, in slow fading, the precoding selection
computation need not be carried out so frequently. 

\vspace{-4mm}
\section{Conclusions}
\label{sec5}
\vspace{-1mm}
We presented a simple, single angular parameter based codebook design
for limited feedback precoding in V-BLAST. The proposed precoding 
scheme achieves full-rate for any $N_t$ by design, whereas the 
achievability of full-diversity was established through BER simulations 
for $N_t=N_r=2,3,4$ under ML decoding for 4-QAM/16-QAM. Our simulation 
results showed that in a $2\times 2$ MIMO system, the proposed scheme 
outperformed the Golden code by about 0.6 dB. It performed comparable to 
perfect codes in $3\times 3$ and $4\times 4$ MIMO systems as well. The 
decoding complexity in the proposed scheme was shown to be much less 
compared to that of Golden/perfect codes. It is noted that the feedback 
channel is an additional resource required in precoding schemes. However, 
given that a feedback channel is often available in most practical wireless 
systems (e.g., for adaptive modulation, rate/power control, etc.), and that
the feedback bandwidth required will be very less in slow fading, the 
proposed scheme can be quite attractive for its full-rate, full-diversity,
and low-complexity attributes. Investigations related to applicability of 
the proposed approach to large antenna systems using multiple-parameter 
based precoder designs can be carried out further.


\vspace{-3mm}
{\footnotesize
\begin{thebibliography}{99}
\vspace{-1.0mm}
\bibitem{paulraj}
A. Paulraj, R. Nabar, and D. Gore, {\em Introduction to Space-Time Wireless
Communications}, Cambridge University Press, 2003.

\bibitem{jafarkhani}
H. Jafarkhani, {\em Space-Time Coding: Theory and Practice}, Cambridge
University Press, 2005.

\bibitem{paul}
H. Bolcskei, D. Gesbert, C. B. Papadias, ans Alle-Jan van der Veen,
Ed., {\em Space-Time Wireless Systems: From Array Processing to
MIMO Communications,} Cambridge University Press, 2006.

\bibitem{bsr}
B. A. Sethuraman, B. S. Rajan, and V. Shashidhar, ``Full-diversity
high-rate space-time block codes from division algebras,'' {\em IEEE
Trans. Inform. Theory}, vol. 49, no. 10, pp. 2596-2616, October 2003.

\bibitem{gold05}
J.-C. Belfiore, G. Rekaya, and E. Viterbo, ``The golden code: A $2\times 2$
full-rate space-time code with non-vanishing determinants,'' {\em IEEE
Trans. Inform. Theory}, vol. 51, no. 4, pp. 1432-1436, April 2005.

\bibitem{OGG}
F. Oggier, J. C. Belfiore, and E. Viterbo, {\em Cyclic Divisional Algebras: 
A Tool for Space-Time Coding}, Foundations and Trends in Communications and 
Information Theory, vol. 4, no. 1, (2007) 1-95.

\bibitem{perf06}
F. E. Oggier, G. Rekaya, J.-C. Belfiore, and E. Viterbo, ``Perfect
space-time block codes,'' {\em IEEE Trans. on Inform. Theory},
vol. 52, no. 9, September 2006.

\bibitem{perf07}
P. Elia, B. A. Sethuraman, and P. V. Kumar, ``Perfect space-time codes for
any number of antennas,'' {\em IEEE Trans. Inform. Theory,}
vol. 53, no. 11, pp. 3853-3868, November 2007.

\bibitem{prec1}
A. Scaglione, P. Stoica, S. Barbarossa, G. B. Giannakis, and H. Sampath,
``Optimal designs for space-time linear precoders and decoders,'' 
{\em IEEE Trans. Signal Process.,} vol. 50, no. 5, pp. 1051–1064, May 2002.

\bibitem{prec2}
H. Sampath and A. Paulraj, ``Linear precoding for space-time coded systems
with known fading correlations,'' {\em IEEE Commun. Lett}., vol. 6, no. 6, 
pp. 239–241, June 2002.

\bibitem{collin}
L. Collin, O. Berder, P. Rostaing, and G. Burel, ``Optimal minimum 
distance-based precoder for MIMO spatial multiplexing systems,'' {\em IEEE
Trans. Signal Process.,} vol. 52, no. 3, pp. 617–627, Mar. 2004.

\bibitem{ucd}
Y. Jiang, J. Li, and W. W. Hager, ``Uniform channel decomposition
for MIMO communications,'' {\em IEEE Trans. Sig. Proc.,} vol. 53, no. 11, 
pp. 4283-4294, November 2005.

\bibitem{ASI}
D. J. Love and R. W. Heath, Jr., ``Limited feedback unitary precoding for 
spatial multiplexing systems,'' {\em IEEE Trans. on Inform. Theory}, 
vol. 51, No. 8, 2967-2976, 2005. 

\bibitem{SL}
C. Simon and G. Leus, ``Feedback quantization for linear precoded spatial 
multiplexing,'' {\em EURASIP Jl. on Advances in Sig. Proc.}, vol. 2008, 
Article ID 683030.

\bibitem{sumeet}
R. W. Heath, Jr., S. Sandhu, and A. Paulraj, ``Antenna selection for 
spatial multiplexing systems with linear receivers,'' {\em IEEE Commun. 
Lett.}, vol.  5, no. 4, pp. 142–144, April 2001.

\bibitem{KAU}
Y. Wang and K. Roy, ``A new reduced complexity sphere decoder with true 
lattice boundary awareness for multi-antenna systems,'' {\em Proc. ISCAS}, 
vol. 5, pp. 4963-4966, May 2005.

\bibitem{VAR}
E.Agrell, T. Eriksson, A. Vardy and K. Zeger, ``Closest point search in 
lattices,'' {\em IEEE Trans. Inform. Theory}, vol. 48, No. 8, 2002.

\end{thebibliography}			 
}
\end{document}